\documentclass{mn2e}
\usepackage{amssymb}
\usepackage{times}
\usepackage{ulem}
\usepackage{color}
\input epsf.sty
\newif\ifAMStwofonts
\AMStwofontstrue

\newlength{\veron}
\newlength{\hhh}
\newcommand{\hh}{\hspace{\hhh}}
\newlength{\mmm}
\newcommand{\mm}{\hspace{\mmm}}
\newlength{\ooo}
\newcommand{\oo}{\hspace{\ooo}}
\newlength{\amm}
\newcommand{\am}{\hspace{\amm}}
\newlength{\ass}
\newcommand{\as}{\hspace{\ass}}
\def\mbh{$M_{\rm BH}$}
\def\mbhlit{M_{\rm BH,\,L}}
\def\mbhvar{M_{\rm BH, \sigma^2}}

\def\Msun{$\rm M_{\odot}$}
\def\Rblr{R_{\rm BLR}}
\def\vblr{v_{\rm BLR}}
\def\hfb{$\nu_{\rm br}$}
\def\Thfb{$T_{\rm br}$}

\def\rms{$\sigma^2_{\rm nxs}$}

\def\tr{\widetilde{r}}

\def\mdot{\dot m}
\def\fwhm{FWHM(H{\small$\beta$})}
\newcommand{\Gs}{$\Gamma_{\rm 0.1-2.4 keV}$}
\newcommand{\Gh}{$\Gamma_{\rm 2-10 keV}$}
\def\kms{kms$^{-1}$}
\newcommand{\lam}{$\lambda$}
\newcommand{\Hb}{H{\small$\beta$}}
\newcommand{\FeII}{Fe\,{\small II}}
\newcommand{\OIII}{[O\,{\small III}]}
\begin{document}
 
\title[Black hole masses in NLS1 galaxies]
%{Black hole masses in NLS1 galaxies. A new type of NLS1?}
{NLS1 galaxies and estimation of their central black hole masses
from the X-ray excess variance method} 

\author[Niko\l ajuk, Czerny \& Gurynowicz]
  {M.~Niko\l ajuk$^1$, B.~Czerny$^2$  and P. Gurynowicz$^1$\\
   $^1$Faculty of Physics, University of Bia\l ystok,
    Lipowa 41, 15-424 Bia\l ystok, Poland \\ 
   $^2$Copernicus Astronomical Center, Bartycka 18, 00-716 Warsaw, Poland}

\maketitle
\begin{abstract}
Black hole mass determination in active galaxies is a key issue in
understanding various luminosity states. In the present paper we try
to generalise the mass determination method based on the X-ray excess
variance, successfully used for typical broad line Seyfert~1 galaxies
(BLS1) to Narrow Line Seyfert~1 (NLS1) galaxies. NLS1 galaxies differ
from BLS1 with respect to several properties. They are generally more
variable in 2-10~ keV energy band so the natural expectation is the
need to use a different scaling coefficient between the mass and the
variance in these two types of sources. However, we find that such a
simple approach is not enough. Although for majority of the 21 NLS1
galaxies in our sample a single scaling coefficient (larger by a
factor 20) provided us with a satisfactory method of mass
determination, in a small subset of NLS1 galaxies this approach
failed. Variability of those objects appeared to be at the
intermediate level between NLS1 and BLS1 galaxies. These exceptional
NLS1 galaxies have much harder soft X-ray spectra than majority of
NLS1 galaxies. We thus postulate that the division of Seyfert 1
galaxies into BLS1 and NLS1 according to the widths of the H$\beta$
line is less generic than according to the soft X-ray slope.
\end{abstract}


%%%%%%%%%%%%%%%%%%%%%%%%%%%%%%%%%%%%%%%%%%%%%%%%%%%%%%%%%%%%%%%%%%%%%%%
\section{Introduction}

Active Galactic Nuclei (AGN) are strongly variable in X-ray band. In a
fixed time-scale quasars containing very massive black holes vary less
than much less massive Seyfert galaxies so the X-ray variability was
used to estimate the black hole mass (e.g. Hayashida et al. 1998; Lu
\& Yu 2001, Markowitz \& Edelson 2004, Papadakis 2004).  However, if
we aim to have a mass determination method which works relatively
accurately we must take into account that the black hole mass is not
the only parameter of the accreting system. The second, important
parameter is the dimensionless accretion rate which may affect the
variability.

In the literature, the three possibilities were suggested for the
exact scaling of the X-ray variability of an AGN with mass and the
accretion rate.

The first and the oldest one is the claim that the variability scales
only with the X-ray luminosity (Barr \& Mushotzky 1986, Lawrence \&
Papadakis 1993, Green, McHardy \& Lehto 1993, Nandra et al. 1997,
Leighly 1999a, Markowitz \& Edelson 2004).  It was recently argued by
Liu \& Zhang (2008) at the basis of a sample of 14 objects with black
hole mass obtained directly from reverberation that the mass itself
does not affect the variations.

The second possibility is a continuous two-parameter relation between
the X-ray variability and the black hole mass and the bolometric (not
X-ray !) luminosity found by McHardy et al. (2006). About this idea
see also papers of Markowitz et al. 2003, Bian \& Zhao 2003, McHardy
et al. 2005, O'Neill et al. 2005.

The third possibility is that there is no continuous scaling with
accretion rate but instead there are two separate scaling laws with
mass: one for Narrow Line Seyfert 1 (NLS1) galaxies and another one
for Broad Line Seyfert 1 (BLS1) galaxies (Hayashida et al. 1998,
Turner et al. 1999, Lu \& Yu 2001, Uttley, McHardy \& Papadakis 2002,
Nikolajuk, Papadakis \& Czerny 2004). This possibility is an
interesting option since the two classes of objects differ
considerably.

The Narrow Line Seyfert~1 (NLS1) galaxies were introduced as a
separate class by Osterbrock \& Pogge (1985). These objects had all
properties in the optical band of the Seyfert 1 galaxies but their
line widths were surprisingly narrow in comparison to typical Seyfert
1s. The original division of Seyfert galaxies, according to the width,
into narrow and broad line objects (Seyfert~1 and Seyfert~2) seemed to
be accompanied by the increase of the \OIII\lam5007 to the \Hb\ flux
ratios (Shuder \& Osterbrock 1981). However, a class of objects with
narrow lines but faint \OIII\lam5007 was found and a new class had to
be introduces.

Till now the NLS1 as a class are defined by the requirement that the 
\OIII\lam5007 to the \Hb\ flux ratios are smaller than 3 and Full Width 
Half Maximum (FWHM) of \Hb\ line in NLS1s was less than 2000
\kms. Further observations revealed other peculiar properties. Blue
wings of some emission lines were stronger compared to that in Seyfert
1s. Also the \FeII\ lines are more asymmetric and much stronger than
in Seyfert 1s (Stephens 1989; Boroson \& Green 1992; Boller et
al. 1996; Wang et al. 1996; Leighly 1999b; V{\'e}ron-Cetty \&
V{\'e}ron 2000; Grupe et al. 2004). The Baldwin effect, which is
common in BLS1s, was seen in those objects as well (e.g. Warner et
al. 2004; Shields 2007). However, Leighly \& Moore (2004) observed the
UV lines and found that the Equivalent Widths of emission lines in
NLS1s were systematically offset to lower values at a given continuum
luminosity in comparison to BLS1 galaxies.

Further differences came from X-ray observations. NLS1s very
frequently exhibit rapid and/or high-amplitude X-ray variability. They
systematically show larger amplitude variations than BLS1 on
time-scales of minutes, hours and days (Hayashida et al. 1998; Fiore
et al. 1998; Leighly 1999a; Turner et al. 1999; Markowitz \& Edelson
2004; Uttley \& McHardy 2005).  Giant-amplitude X-ray variability has
been observed in several NLS1s (e.g. in NGC 4051 - Uttley et al. 1999;
in PHL 1092 - Brandt et al. 1999).
%We would like to note that the same behaviour in larger amplitude 
%of X-ray has been detected in a few BLS1s, as well.  
%For instance, NGC 3227 displayed X-ray flux variability 
%by a factor of fifteen (Kommosa \& Fink 1997a; Lamer, Uttley 
%\& McHardy 2003), NGC 3786 by a factor of ten (Kommosa \& Fink 1997b).
The X-ray spectra of NLS1s are generally steeper than spectra of BLS1s
(Brandt, Mathur \& Elvis 1997). The average photon index, \Gh\, in
BLS1 in 2-10 keV band is $\simeq 1.7$ 
%\Gh $=1.73 \pm 0.05$ 
(Nandra \& Pounds 1994; Reynolds 1997), whereas NLS1 galaxies have
\Gh\ $\simeq 2.2$ (Leighly 1999a). 
%\Gh $=2.19$ 
Similar effect is in 17-100 keV band. The photons index in BLS1
galaxies is $\Gamma_{\rm 17-100 keV} = 2.02 \pm 0.16$ and in NLS1
$\Gamma_{\rm 17-100 keV} = 2.6 \pm 0.3$ (Malizia et al. 2008).  The
effect is much stronger in the soft X-ray band, 0.1-2.4 keV. The
photon index, \Gs, of BLS1s varies around value 2.1-2.3 (Boller et
al. 1996; Grupe et al. 1998; Pfefferkorn et al. 2001) but in NLS1s it
is typically of order of 3.0-3.3, and occasionally even larger (Boller
et al. 1996; Lawrence et al. 1997; V{\'e}ron-Cetty et al. 2001;
Williams et al. 2004; Bian 2005). More precisely, NLS1s may show both
steep as well as flat X-ray spectra, while BLS1s always have flat
spectra (Grupe et al. 1999).  Generally, there is an overall
anticorrelation between the soft photon spectral index \Gs\ and the
Balmer line width \fwhm\ (Puchnarewicz et al. 1992; Boller et
al. 1996; Brandt et al. 1997; Williams et al. 2004). Recently Zhou et
al. (2006) have reexamined this relationship using larger sample of
objects and concluded that the anticorrelation in NLS1s extends only
to FWHM $\sim$ 1000 \kms, and at smaller line widths the trend appears
to reverse.

Interpretation of observational data suggests that NLS1s are Seyfert
1s, which have relatively low mass for a given luminosity and
therefore radiate near the Eddington limit (Pounds et al. 1995;
Hayashida 2000; Shrader \& Titarchuk 2003; Botte et al. 2004); even
super-Eddington accretion rates were suggested (Wang et al. 1996;
Collin \& Kawaguchi 2004; Zhang \& Wang 2006). Estimated Eddington
ratios as well as the spectral properties may indicate that NLS1s
represent the supermassive BH analogue of Galactic BH (GBH) in their
high states (Pounds et al. 1995; McHardy et al. 2004; Gierli\' nski \&
Done 2006; Sobolewska \& Done 2007). However, more research should be
done to confirm --- or reject --- this view, and thus the black hole
mass determination is one of the key issues.

In the present paper we determine the X-ray excess variance for a
larger sample of 21 NLS1 galaxies and discuss the scaling properties
in the combined sample of NLS1 and BLS1 objects.

The methods of mass determination are given in Sect.~\ref{sect:method},
the obtained results are shown in Sect.~\ref{sect:results}, and we
discuss the results in Sect.~\ref{sect:discussion}.

%<<<<<<<<<<<<<<<<<<<<<<<<<<<<<<<<<<<<<<<<<<<<<<<<<<<<<<<<<<<<<<<<<<<<<<
\section{Method}
\label{sect:method}

%\subsection{Selected sample of Seyfert galaxies}
%\label{sect:sample}

We select a sample of NLS1 galaxies based on the requirement that an
object was well studied both in optical and X-ray band. Majority of
the classical NLS1 sources have the low redshift values (z $<0.1$) and
low X-ray luminosities in 2-10 keV band. Additionally we add two BLS1s
-- dwarf active galaxies which are close to being Sy2. First object is
the least luminous AGN known - NGC 4395, and the second object is also
a low luminosity source, with jet, and with maser emission - NGC
4258. Table~\ref{tab:tab1} shows our sample of 23 selected sources.
Types, redshifts and coordinates of the galaxies were taken from a
catalogue of quasars and active nuclei (V{\'e}ron-Cetty \& V{\'e}ron,
2006).

We supplement it with the sample of BLS1 galaxies
taken from Nikolajuk et al. 2006 (see Table~\ref{tab:tab2}).

%%%%%%%%%%%%%%%%%%%%%%%%%%%%%%%%%%%%%%%%%%%%%%%%%%%%%%%%%%%%%%%%%%%%%%%
%%%%%%%%%%%%%%%%%%%%%%%%%%%%%%%%%%%%%%%%%%%%%%%%%%%%%%%%%%%%%%%%%%%%%%%
\begin{table*}
\caption{The NLS1 sources in our sample supplemented with two BLS1 objects.}
\label{tab:tab1}
\begin{center}
\begin{tabular}[t]{lcccccccc}
\hline
Name & $\rm RA(J2000.0)$&$\rm DEC(J2000.0)$&z&Type&FWHM($\rm H\beta$)& ref.&$\Gamma_{\rm 0.1\div 2.4 keV}$& ref.\\
\hline
Mrk 335        &$00^h06^m19.5^s$ &$+20^{\circ}12'10''$&0.026& NLS1     &$1710\pm140$&2& $3.10\pm 0.05$ & 1\\
I Zw 1         &$00\hh53\mm34.9$ &$+12\oo41\am36\as$  &0.061& NLS1/QSO & 1240       &1& $3.09\pm 0.16$ & 1\\
TON S180       &$00\hh57\mm19.9$ &$-22\oo22\am59\as$  &0.062& NLS1/QSO & $970\pm100$&2& $3.04\pm 0.01$ & 1\\
PHL 1092       &$01\hh39\mm55.8$ &$+06\oo19\am21\as$  &0.396& NLS1/QSO & 1790       &1& $4.3 \pm 0.3 $ & 1\\
1H 0707-495    &$07\hh08\mm41.5$ &$-49\oo33\am06\as$  &0.041& NLS1     & 1050       &1& $2.25\pm 0.25$ & 1\\
Mrk 110        &$09\hh25\mm12.9$ &$+52\oo17\am11\as$  &0.036& NLS1     &$1760\pm50$ &2& $2.17\pm 0.27$ & 9\\
Mrk 142        &$10\hh25\mm31.3$ &$+51\oo40\am35\as$  &0.045& NLS1     &$1620\pm120$&2& $3.15\pm 0.12$ & 1\\
Mrk 42         &$11\hh53\mm41.8$ &$+46\oo12\am43\as$  &0.025& NLS1     &670-865   &3,4& $2.6 \pm 0.2 $ & 8\\
NGC 4051       &$12\hh03\mm09.6$ &$+44\oo31\am53\as$  &0.002& NLS1     &$1170\pm100$&2& $2.84\pm 0.04$ & 1\\
NGC 4395       &$12\hh13\mm48.9$ &$+33\oo42\am48\as$  &0.001& BLS1/S1.8& 1500	    &5& $\triangle$	&\\
PG 1211+143    &$12\hh14\mm17.7$ &$+14\oo03\am13\as$  &0.081& NLS1/QSO &$1900\pm150$&2& $3.03\pm 0.15$ & 1\\
Mrk 766        &$12\hh18\mm26.5$ &$+29\oo48\am46\as$  &0.012& NLS1     &$1100\pm200$&2& $2.75\pm 0.13$ & 9\\
%Gamma=1.68\pm0.03 Branduardi-Raymont et al. (2001)
%Gamma=2.79\pm0.11 (1); 2.5\pm0.2 (8)
NGC 4258       &$12\hh18\mm57.5$ &$+47\oo18\am14\as$  &0.001& BLS1/S1.9&$1200\pm400$ $\star$&6&$\Box$& \\
%B.J.Wilkes,G.D.Schmidt,P.S.Smith,S.Mathur,K.K. McLeod,1996,ApJ,455,L13 => FWHMHbeta(polarized)=1100 km/s
PG 1244+026    &$12\hh46\mm35.2$ &$+02\oo22\am09\as$  &0.048& NLS1     &$830\pm50$&2& $3.26\pm 0.14$ & 1\\
MCG -6-30-15   &$13\hh35\mm53.8$ &$-34\oo17\am44\as$  &0.008& NLS1     &$1700\pm170$&7& $2.33\pm 0.23$ &9\\
%Gamma=1.33\pm0.01 Branduardi-Raymont et al. (2001)
PG 1404+226    &$14\hh06\mm21.8$ &$+22\oo23\am46\as$  &0.098& NLS1     &880-950&1,4& $4.3 \pm 0.3 $ & 1\\
NGC 5506       &$14\hh13\mm14.8$ &$-03\oo12\am27\as$  &0.006& NLS1/S2.0&     &     &                &  \\
Mrk 478        &$14\hh42\mm07.4$ &$+35\oo26\am23\as$  &0.079& NLS1/QSO &$1630\pm150$&2& $3.06\pm 0.03$ & 1\\
HB89 1557+272  &$15\hh59\mm22.2$ &$+27\oo03\am39\as$  &0.065& NLS1     & 1410 &8& $1.3 \pm 0.6 $ & 3\\
IRAS 17020+4544&$17\hh03\mm30.4$ &$+45\oo40\am47\as$  &0.060& NLS1     & 1040&1    & $2.37\pm 0.19$ & 1\\
Mrk 507        &$17\hh48\mm38.4$ &$+68\oo42\am16\as$  &0.056& NLS1     &960-1565&8,4& $1.68\pm 0.16$ & 1\\
IC 5063        &$20\hh52\mm02.3$ &$-57\oo04\am08\as$  &0.011& NLS1/S2.0&     &     &                &  \\
Akn 564        &$22\hh42\mm39.3$ &$+29\oo43\am31\as$  &0.025& NLS1     &720-950&8,1& $3.47\pm 0.07$ & 1\\
\hline
\end{tabular}
\end{center}
Column~(1)-(4) list the object name, coordinates and redshifts. Column~(5)
list types of the galaxies. NLS1 or BLS1 mean the overall type of the
galaxy. QSO (in the second part, after /) denotes that this source is
brighter in B colour than absolute magnitude -23 mag ($M_{\rm B} <
-23$). In other case of remaining NLS1s or BLS1s we have galaxies that
are fainter than -23 mag.  Columns (6) and (8) list the values of the
FWHM of \Hb\ (broad component) in \kms\ and the values of the soft
photon index, \Gs, respectively. Both values were taken from
literature. For NGC~4258, the sign $\star$ denotes FWHM of polarized
\Hb.  The sign $\triangle$ in Column~(8) denotes that $\Gamma$ (above
1~keV) varies significantly on time-scales of 1 year or less ($\Gamma
\sim$ 0.6-1.72; Moran et al. 2005). The~$\Box$~indicates that the
nuclear component is visible above 2~keV only (Fiore et al. 2001). The
numbers in Column~(7) and (9) correspond to the following references: 1
- Leighly (1999b), 2 - Grupe et al. (2004), 3 - Bian \& Zhao
(2003), 4 - Wang \& Lu (2001), 5 - Kraemer et al. (1999), 6 - Barth
et al. (1999), 7 - Nandra (2006), 8 - Boller et al. (1996), 9 - Walter
\& Fink (1993).
\end{table*}
%%%%%%%%%%%%%%%%%%%%%%%%%%%%%%%%%%%%%%%%%%%%%%%%%%%%%%%%%%%%%%%%%%%%%%%
%%%%%%%%%%%%%%%%%%%%%%%%%%%%%%%%%%%%%%%%%%%%%%%%%%%%%%%%%%%%%%%%%%%%%%%
\begin{table}
\caption{The BLS1 sources in our sample taken from Nikolajuk et al. 2006.}
\label{tab:tab2}
\begin{center}
\begin{tabular}[b]{lcccc}
\hline
Name & FWHM($\rm H\beta$)& ref.&$\Gamma_{\rm 0.1\div 2.4 keV}$& ref.\\
\hline
3C 120		&$2360\pm 170$&1&	& \\
3C 390.3	&$9630\pm 804$&2& $1.87\pm 0.45$ & 3\\
Ark 120		&$5536\pm 297$&2& $2.63\pm 0.29$ & 3\\
IC 4329A	&$5620\pm 200$&1& $1.71\pm 0.87$ & 3\\
Mrk 509		&$3430\pm 240$&1& $2.34\pm 0.16$ & 3\\
NGC 3227	&$5138\pm 787$&2&	& \\
NGC 3516	&$3353\pm 310$&1& 	& \\
NGC 3783	&$3570\pm 190$&1& $3.16\pm 0.63$ & 3\\
NGC 4151	&$4248\pm 516$&2&	& \\
NGC 4593	&$5320\pm 610$&1& $2.49\pm 0.42$ & 3\\
NGC 5548	&$5830\pm 230$&1& $2.21\pm 0.15$ & 3\\
NGC 7469	&$2650\pm 220$&1& $2.52\pm 0.24$ & 3\\
F 9		&$6270\pm 290$&1& $2.21\pm 0.19$ & 3\\
\hline
\end{tabular}
\end{center}
Column~(1) list the object name.  Columns (2) and (4) list the values of
the FWHM of \Hb\ (broad component) in \kms\ and the values of the soft
photon index, \Gs, respectively. The numbers in Columns~(3) and (5)
correspond to the following references: 1 - Nandra (2006), 2 -
Peterson et al. (2004), 3 - Walter \& Fink (1993).
\end{table}
%%%%%%%%%%%%%%%%%%%%%%%%%%%%%%%%%%%%%%%%%%%%%%%%%%%%%%%%%%%%%%%%%%%%%%%

%----------------------------------------------------------------------
\subsection{Black hole mass from method based on optical observations}

The estimation of the black hole masses is difficult.  Several methods
were developed for that purpose. We tried to collect the black hole
mass determinations for our sample in possibly consistent way.

The most precise method to determine \mbh\ is based on water maser
emission and the black hole mass from this method was used for NGC
4258.

Most of the sources were subject of AGN monitoring campaign so the
reverberation method could have been used (e.g. Blandford \& McKee
1982; Peterson \& Wandel 1999, 2000; Vestergaard 2002). This technique
assumes that the motion of the gas moving around the black hole is
dominated by the gravitational force and therefore the gaseous clouds
in Broad Line Region (BLR) are virialized. Hence the central black
hole mass can be estimated using the BLR radius, $\Rblr$, and velocity
of the BLR gas, $\vblr$: $M_{\rm BH} = \Rblr \vblr^2/G$, where G is
the gravitational constant.  The radius $\Rblr$ can be estimated from
observations of a light travel time $\tau$. One can assume that an
optical/UV continuum, which is produced in an accretion disc,
photoionises BLR clouds and leads to formation of emission lines
(e.g. \Hb). The continuum varies in time and therefore the lines in
BLR also vary but they respond with a delay $\tau$. The value of
$\tau$ can be thus interpreted as a light travel distance $\Rblr$.
Such an approach to estimations of $\Rblr$ distance is accurate, and
it is later referred as 'reverberation method'. However, it requires a
long observational campaign, which are time consuming. In order to
save that time, the radius $\Rblr$ can be estimated from the empirical
relationship between the size and the monochromatic luminosity at
5100\AA\ (Kaspi et al. 2000):
\begin{eqnarray}
\label{eq:Rblr}
\Rblr = (32.9^{+2.0}_{-1.9}) \Big[ 
\frac{\lambda L_{\lambda}(5100 \rm \AA)}{10^{44}  {\rm erg s^{-1}}} 
\Big]^{0.700\pm0.033} \quad {\rm light-days}\ .
\end{eqnarray}
Such approach is called 'scaling method' and it was used by many
scientist (Wandel 1999; Kaspi et al. 2000, 2005; Bentz et al. 2006;
McGill et al. 2008). As to the velocity dispersion, it can be
expressed as $\vblr = f \times$FWHM, where FWHM is the width of
e.g. \Hb\ line and the geometrical factor $f$ depends on the geometry
and the details of the gas kinematics in BLR (Krolik 2001; McLure \&
Dunlop 2001; Collin et al. 2006). Many authors assume the isotropy of
the gas motions in BLR. In such case the factor $f$ is equal to
$\sqrt{3}/2$ (Netzer 1990, Kaspi et al. 2000) and the velocity can be
written as $\vblr = (\sqrt{3}/2)$\fwhm. Onken et al. (2004) determined
new value of the factor $f$.  They have used the correlation between
the black hole mass and the stellar velocity dispersion in bulge
mainly in BLS1 galaxies. They found out $f \simeq \sqrt{5.5}/2 =
1.17$. Such value of $f$ was adopted by Peterson et al. (2004) and
Peterson et al. (2005) in the estimation of BH masses in the
reverberation method. We adopt those mass measurements when available
(i.e. for sources Mrk 335, Mrk 110, NGC 4051, NGC 4395 and
PG1212+143). Several other objects (I Zw1, TON S180, PHL 1092, 1H
0707, Mrk 142, Mrk 42, Mrk 766, PG 1244, MCG -6-30-15, PG 1404, Mrk
478, IRAS 17020, Mrk 507, Ark 564) had mass determination based on the
old $f$ coefficient used by Kaspi and those masses were corrected for
the two coefficients ratio, i.e. by the factor $\frac{4}{3}
\frac{5.5}{4} = 1.83$ in order to obtain $\mbhlit$ values. The
original values and the rescaled values are both given in
Table~\ref{tab:Optdata}. When we have a choice between the black hole
masses from Wang \& Lu (2001) or Bian \& Zhao (2003) (both masses
estimated through the scaling method based on relationship between BLR
radius and optical luminosity at 5100~\AA) we have decided to take
$\mbhlit$ from the later paper. We believe that masses estimated in
2003 are determined more precisely that in 2001.

For two objects (NGC~5506 and in IC~5063) the black hole mass was
estimated from stellar velocity dispersion, and the determination of
the mass for HB89 1557+272 is outlined below.

All mass values and the references are collected in
Table~\ref{tab:Optdata}.

%%%%%%%%%%%%%%%%%%%%%%%%%%%%%%%%%%%%%%%%%%%%%%%%%%%%%%%%%%%%%%%%%%%%%%
\begin{table}
\caption{The black hole masses taken from literature and/or
corrected by us for NLS1 galaxies and two BLS1 objects.}
\label{tab:Optdata}
\begin{center}
\begin{tabular}{l c c c}%
\hline
Name & $M^{\rm old}$ & $\mbhlit$ & ref./ \\
     &($10^{6}\ \rm M_{\odot}$)&($10^{6}\ \rm M_{\odot}$)& meth. \\
\hline
Mrk 335        & 			&$14.2\pm3.7$		& 1/r\\
I Zw 1         &$18.20^{+5.24}_{-4.07}$	&$33.4^{+9.6}_{-7.5}$	& 2/s\\
TON S180       &$11.6$-$12.3$ 	 	&21.3-22.5		& 3,5/s\\
PHL 1092       &$160\pm60$ 		&$293\pm110$   		& 4/s\\
1H 0707+495    &$2.04$ 			& 3.74			& 5/s\\
Mrk 110        & 			&$25.1\pm6.1$		& 1/r\\
Mrk 142        & 4.7 			& 8.62			& 3/s\\
Mrk 42         &$0.4$-$10.0$ 		&$0.73$-$18.3$		& 5,3/s\\
NGC 4051       & 			&$1.91\pm0.78$ 		& 1/r\\
NGC 4395       & 			&$0.36^{+0.11}_{-0.11}$ & 6/r\\
PG 1211+143    & 			&$146\pm44$    		& 1/r\\
Mrk 766        &$0.83$-$4.3$ 		&$1.52$-$7.88$		& 5,3/s\\
NGC 4258       &			&$39\pm1$		& 7/m\\
PG 1244+026    & 1.3 			& 2.38  		& 3/s\\
MCG -6-30-15   &$1.55\pm0.30$ 		&$2.84\pm0.55$		& 5/s\\
PG 1404+226    &$6.6$-$10.0$ 		&$12.1$-$18.3$		& 5,3/s\\
NGC 5506       &			&$88$			& 8/d\\
Mrk 478        &$18.7$-$21.9$ 		&$34.3$-$40.2$		& 3,5/s\\
HB89 1557+272  &			& 4.17 			& \\
IRAS 17020+45  & 5.9 			& 10.8			& 3/s\\
Mrk 507        &11.6 			& 21.3			& 3/s\\
IC 5063        &			& 55    		& 9/d\\
Ark 564        &$1.2$-$2.9$ 		&$2.20$-$5.32$		& 5,3/s\\
\hline
\end{tabular}
\end{center}
Column~(1) lists the object name. Column~(2) shows the masses of black
holes obtained by using the scaling method (based on relationship
between BLR radius and optical luminosity).  Those values $M^{\rm
old}$ have been corrected by us and shown in Column~(3) (see
text). Column~(3) also lists \mbh\ values of those sources for which
masses were obtained from other methods (see the letter in Column~(4)).
In case of HB89 1557+272 their \mbh\ was calculated (see text for
details). Values $\mbhlit$ were taken for further analysis. The
numbers in Column~(4) correspond to the following references: 1 -
Peterson et al. (2004), 2 - Vestergaard (2002), 3 - Wang \& Lu (2001),
4 - Dasgupta et al. (2004), 5 - Bian \& Zhao (2003), 6 - Peterson et
al. (2005), 7 - Herrnstein et al. (1998), 8 - Papadakis (2004) 9 - Woo
\& Urry (2002).  The letter in Column~(4) shows the method of
estimations of \mbh: 'r' - the reverberation mapping technique, 's' -
the scaling method, 'd' - stellar velocity dispersion, 'm' - maser).
\end{table}

%-----------------------------------------------------------------
\subsubsection{The case of HB89 1557+272}

No black hole mass estimate for HB89 1557+272 can be found in the
literature, and the object was not monitored.  Therefore, we obtain a
simple order of magnitude estimate in a following way.  From Malkan,
Margon \& Chanan (1984) we take the total bulge magnitude in r
(6650\AA; defined by e.g. Thuan \& Gunn 1976) colour. This value
$m_{\rm r,bulge}$ is equal to 17.05. Then we convert this luminosity
to the bulge magnitude in B (4400\AA) colour.  For this purpose, we
assume that the bulge spectrum in HB89 1557+272 is the same as in
M31. Using it (see Wamsteker et al. 1990) we obtain the ratio
$F_{\nu}(6650\rm\AA)/F_{\nu}(4400\rm\AA)= 1.24$. We convert $m_{\rm
r,bulge}$ into $m_{\rm B,bulge}$ based on this information and
formulas from Zombeck (1990). We obtain that $m_{\rm B,bulge} = 17.73$
mag. In the third step, we calculate $L_{\rm B, bulge}$.  Due to small
redshift of this object (z=0.064625) we assume that in our universe
the cosmological constant $\Lambda = 0$.  The Hubble constant and the
deceleration parameter are $H_0$ = 75 km s$^{-1}$ Mpc$^{-1}$ and $q =
0.5$, respectively.  Under those conditions we obtain the bulge
luminosity in the B colour $L_{\rm B,bulge} = 1.28 \times 10^{43}$ erg
s$^{-1}$. Kormedy \& Gebhardt (2001) have presented the correlation
between \mbh\ and $L_{\rm B,bulge}$:
\begin{eqnarray}
\label{eq:Lbulge}
M_{\rm BH} = 0.78 \times 10^8 \Big( 
\frac{L_{\rm B,bulge}}{10^{10} L_{\rm B,\odot}}\Big)^{1.08} \ ,
\end{eqnarray}
where $L_{\rm B,\odot}$ denotes the Sun luminosity in B band\footnote{
$L_{\rm B,\odot} = 1.93\times10^{33}$ erg s$^{-1}$}. Hence the black
hole mass obtained from the formula is equal to $5.00 \times 10^7
M_{\odot}$. Kormedy \& Gebhardt's relationship is worth only for
normal galaxies and BLS1.  V{\'e}ron-Cetty \& V{\'e}ron (2006)
classify HB89 1557+272 as ordinary NLS1 sources, because it luminosity
in B band is low.  However, Malkan et al. (1984), Hutching et al. (1984)
or NED\footnote{NASA/IPAC Extragalactic Database} classify this object
as quasar with the narrow emission lines. We assume the narrow line type
of HB89 1557+272 and have to correct obtained \mbh\ in order to
estimate true value of it.  Mathur, Kuraszkiewicz \& Czerny (2001)
have mentioned that the BH to bulge mass ratio for NLS1 galaxies is of
order of 0.00005 and 0.0005 for NL quasars. For Wandel's (1999) sample
of galaxies this ratio \mbh/$M_{\rm bulge}$ = 0.0003 for BLS1 and 0.006
for normal galaxies and quasars. We adopt those values in our
analysis. If we assume that (i) the bulge masses in NLS1 and BLS1
galaxies are the same ($M_{\rm bulge}^{\rm NLS1} = M_{\rm bulge}^{\rm
BLS1}$), (ii) the BH masses in NLS1s are simultaneously smaller
than in BLS1s ($M_{\rm BH}^{\rm NLS1} < M_{\rm BH}^{\rm BLS1}$),
and (iii) there is only a systematic shift in
equation~(\ref{eq:Lbulge}) between those two kinds of Seyfert 1
galaxies, we obtain that $M_{\rm BH}^{\rm BLS1} = 6\times M_{\rm
BH}^{\rm NLS1}$ and $M_{\rm BH}^{\rm BLS1/QSO} = 12\times M_{\rm
BH}^{\rm NLS1/QSO}$. Therefore, we should divide the value
$5.00\times10^7$ by 6 in the case if HB89 1557+272 would be NLS1 and
by 12 if the galaxy would be NL quasars. We choose the last one factor
and the BH mass in HB89 1557+272 is \mbh = $4.17
\times 10^6 M_{\odot}$. We adopted this value in our further analysis 
(Table~\ref{tab:Optdata}). We must note here that the value of \mbh\
calculated by us has a large error and may be underestimated.

%-------------------------------------------------------------------
\subsection{Determination of the X-ray excess variance}
\label{sect:varmeth}

We determine the X-ray excess variance for the sources in our sample
using the approach which was developed by Nikolajuk et al. (2004) with
the aim to use for mass determination in BLS1 galaxies. The approach
is based on the use of only the high frequency tail of the power
spectrum and on the specific way of combining results for several
lightcurves for a given source. 

Each AGN emit a variable X-ray radiation observed in the hard (2-10
keV) band. From such lightcurve we can calculate the normalized excess
variance which is defined as (e.g. Nandra et al. 1997; Turner et
al. 1999):
\begin{eqnarray}
\sigma^2_{\rm nxs} = \frac{1}{N \bar{x}^2} \sum_{i=1}^{N} 
\big[(x_i - \bar{x})^2 - (\delta x_i)^2 \big] \ , 
\label{eq:rms}
\end{eqnarray}
Here, $N$ is the number of data points, $x_i$ and $\delta x_i$ are the
flux and its error, respectively. $\bar{x}$ is the unweighted,
arithmetic mean of $x_i$. \rms\ is in units of (rms/mean)$^2$.
The important value is not the excess variance itself but actually the
ratio of the excess variance to the duration of the lightcurve since
the power spectrum has the slope $\sim 2$ at high frequencies. This is
clearly seen from the formula used to calculate \mbh\ for BLS1
galaxies: 
\begin{eqnarray}
M_{\rm BH} = C \frac{T - 2\Delta t}{\sigma^2_{\rm nxs}}\ ,
\label{eq:Mbh}
\end{eqnarray}
where $T$ is the duration of the X-ray lightcurve (in seconds),
$\Delta t$ is its bin size (in seconds) and $C$ is the constant.  The
requirement of the method is that the excess variance is measured at
the high frequency tail of the power spectrum, i.e. that the length of
a single lightcurve, $T$, is significantly shorter than the inverse of
the high frequency break of the power spectrum, \hfb\ (i.e. $T <
\frac{1}{\nu_{\rm br}}$). The same condition is important for any
scaling relations so we follow it here.  We would like to emphasise
here that we need not to know exact value of $T$. We chose $T$ based
on information that our $T$ should be shorter than break time, \Thfb.
This time, which is simply equal to inversion of $\nu_{\rm br}$
(i.e. \Thfb = $1/\nu_{\rm br}$), is shown in Table~\ref{tab:Xdata}. In
case of those NLS1 sources, for which we do not have any information
about $T_{\rm br}$, we estimate this time in following way. We use the
relationship between the black hole mass and the break frequency in
BLS1 galaxies ($\nu_{\rm br} = 15/(M_{\rm BH}/$\Msun) Hz) derived by
Papadakis (2004). In our calculations we assume $M_{\rm
BH}=\mbhlit$. The break frequencies in NLS1s are higher by a factor
10-30 than in BLS1 sources (e.g. Papadakis 2004). Therefore, we
multiply Papadakis' relationship by a factor $D=10-30$, i.e. our
estimated $T_{\rm br} = \mbhlit/15 D$ s.  Generally, the value $D$ is
equal to 30. Sometimes the calculated excess variance had been less
than zero and in those cases we decreased $D$ in order to obtain
longer $T$. We have decreased $D$ from 30, unless the values of \rms\
become positive, but still $D\ge 10$.  We choose now
$T=[0.7-0.95]\times$\Thfb.

%The value of $D$ depends on
%sign of obtained \rms. In the first step $D=30$. If the excess
%variances are less than zero then we decrease $D$, unless the values
%of \rms become positive, but still $D\ge 10$.  We would like to note
%that we do not require here exact value of \Thfb. Our $T$ only should
%be less than estimated by us \Thfb.

%%%%%%%%%%%%%%%%%%%%%%%%%%%%%%%%%%%%%%%%%%%%%%%%%%%%%%%%%%%%%%%%%%%%%%%%%
\begin{table}
\caption{X-ray observations and selected lightcurves for NLS1 sample 
and two BLS1 objects.}
\label{tab:Xdata}
\begin{tabular}{lclccc}
\hline
Object &$T_{\rm br}$ (ref.)&Proposal&Mean&$N_{\rm lc}$ \\ 
name   & [ks]~~~~~~~~&details&[cnts/s] & \\
\hline
Mrk 335        &&A/71010000 & 0.398 & 6  \\
I Zw 1         &&A/73042000 & 0.149 & 4  \\
TON S180       &&A/77036000 & 0.313 & 34 \\ 
PHL 1092       &&A/75042000 & 0.029 & 4  \\
1H 0707+495    &&A/73043000 & 0.070 & 7  \\
Mrk 110        &&A/73091000 & 0.793 & 3  \\
Mrk 142        &&A/76034000 & 0.120 & 4  \\
               &&A/76034010 & 0.097 & 2  \\
Mrk 42         &&A/75056000 & 0.041 & 4  \\
%no more data from ASCA and RXTE
NGC 4051       &$1.64^{+1.12}_{-0.60}$ (1)&R/P70141 & 2.302 & 10\\
               &&R/P70142   & 2.198 & 1  \\
NGC 4395       &$0.33$-$2.00$ (2)&A/78009000 & 0.098 & 8 \\
               &&A/78009001 & 0.105 & 21 \\				   
               &&A/78009002 & 0.080 & 9  \\				   
               &&A/78009003 & 0.067 & 9  \\
PG 1211+143    &&A/70025000 & 0.171 & 3  \\
Mrk 766        &$2.0^{+3.0}_{-0.4}$  (3)&A/71046000 & 0.469 & 10 \\
%ASCAdata only one
%RXTEdata for Mrk 766 (e.g. 80159;92108;60135) gives nothing -
%too low fluxes, too high errors.
NGC 4258       &$44400^{+\infty}_{-43900}$(4)&A/60018000 & 0.121 & 1 \\
	       &&A/64001000 & 0.176 & 1 \\
	       &&A/64001010 & 0.185 & 1 \\
	       &&A/77031000 & 0.117 & 8 \\
PG 1244+026    &&A/74070000 & 0.162 & 8 \\ 
MCG -6-30-15   &$13.0^{+8.6}_{-7.8}$ (5)&R/P10299 & 4.594 & 1  \\
	       &&R/P20310   & 4.119 & 20 \\
	       &&R/P40155   & 4.346 & 8 \\
	       &&A/70016300 & 1.221 & 3 \\
	       &&A/77003000 & 0.672 & 35 \\			   
PG 1404+226    &&A/72021000 & 0.020 & 2  \\	   
NGC 5506       &$76.9^{+95.0}_{-66.5}$ (6)&R/P20318  & 8.563 & 12\\
Mrk 478        &&A/73067000 & 0.109 & 5  \\
HB89 1557+272  &&A/81004000 & 1.624 & 4  \\
               &&A/81005000 & 0.559 & 6  \\
%R/5015804 gives nothing
IRAS 17020+45  &&A/73047000 & 0.303 & 11 \\
Mrk 507        &&A/74033000 & 0.042 & 1  \\
%no any data for Mrk 507 from RXTE
IC 5063        &&R/P10337   & 0.988 & 1  \\
	       &&A/71030010 & 0.124 & 1 \\	
Ark 564        &$0.59^{+0.66}_{-0.15}$(7)&R/P10291   & 2.035 & 19 \\
\hline
\end{tabular}
Column~(1) lists the object name. Column~(2) shows the power spectral
density (PSD) break time-scales, which are connected with the
frequency break (\Thfb=1/\hfb). The number in parentheses correspond
to the following references: (1) McHardy et al. (2004); (2) Vaughan et
al. (2005); (3) Vaughan \& Fabian (2003); (4) Markowitz \& Uttley
(2005); (5) McHardy et al. (2005); (6) Uttley \& McHardy (2005); (7)
Papadakis et al. (2002). Column~(3) shows the observation details. The
first letter refers to the satellite name (A -- ASCA, R -- RXTE), the
number refers to proposal number. In Column~(4) are shown the mean
fluxes observed in the lightcurve (in count per second units).
Column~(5) lists the number of short lightcurves, $N_{\rm lc}$,
subtracted from whole proposal lightcurve. $N_{\rm lc}$ also means the
number of the values (\rms)$_k$, which were used in order to estimate
$\mbhvar$.
\end{table}

%%%%%%%%%%%%%%%%%%%%%%%%%%%%%%%%%%%%%%%%%%%%%%%%%%%%%%%%%%%%%%%%%%%%%%%%%%

%%%%%%%%%%%%%%%%%%%%%%%%%%%%%%%%%%%%%%%%%%%%%%%%%%%%%%%%%%%%%%%%%%%%%%%
\begin{table}
\caption{The mean time of observations and the mean normalized excess
variances the whole sample of galaxies.}
\label{tab:TsigLx}
\begin{center}
\begin{tabular}{l c c c}%
\hline
Name & $\langle T \rangle$ & $\sigma^2_{\rm nxs}$ &
$L_{2-10 \rm keV}$ \\
	& [s] & ${\rm (rms/mean)^2}$ & \\
\hline
Mrk 335        & 2396 & $4.108^{+2.921}_{-2.259} \times 10^{-3}$ & 43.07 (a)\\
I Zw 1         & 2270 & $1.978^{+2.144}_{-1.708} \times 10^{-3}$ & 43.35 (a)\\
TON S180       & 7842 & $1.194^{+0.253}_{-0.243} \times 10^{-2}$ & 43.58 (a)\\
PHL 1092       &11584 & $3.952^{+3.617}_{-3.720} \times 10^{-2}$ & 44.15 (b) \\
1H 0707+495    & 7696 & $3.512^{+1.994}_{-1.635} \times 10^{-2}$ & 42.49 (b)\\
Mrk 110        & 2632 & $3.476^{+3.217}_{-2.269} \times 10^{-3}$ & 43.80 (c)\\
Mrk 142        & 7510 & $1.625^{+1.073}_{-0.881} \times 10^{-2}$ & 43.17 (a)\\
Mrk 42         & 8162 & $1.173^{+1.593}_{-1.173} \times 10^{-2}$ & 42.12 (b)\\
NGC 4051       & 1500 & $1.498^{+0.449}_{-0.302} \times 10^{-2}$ & 41.21 (a)\\
NGC 4395       & 2100 & $3.085^{+1.183}_{-1.050} \times 10^{-2}$ & 39.99 (a) \\
PG 1211+143    & 7967 & $2.688^{+13.171}_{-2.164} \times 10^{-3}$& 43.63 (b)\\
Mrk 766        & 2460 & $6.326^{+4.645}_{-3.763} \times 10^{-3}$ & 42.73 (a)\\
NGC 4258       &31300 & $6.215^{+1.566}_{-2.350} \times 10^{-3}$ & 40.52 (d) \\
PG 1244+026    & 1200 & $1.459^{+0.922}_{-0.681} \times 10^{-2}$ & 43.03 (a)\\
MCG -6-30-15   & 8408 & $1.065^{+0.211}_{-0.139} \times 10^{-2}$ & 42.72 (a)\\
PG 1404+226    & 7382 & $9.380^{+60.975}_{-5.535} \times 10^{-3}$& 43.03 (b)\\
NGC 5506       & 8015 & $2.997^{+0.900}_{-0.900} \times 10^{-3}$ & 42.73 (a)\\
Mrk 478        &11153 & $1.265^{+0.964}_{-0.743} \times 10^{-2}$ & 43.50 (a)\\
HB89 1557+272  & 2354 & $2.269^{+1.106}_{-1.308} \times 10^{-3}$ & \\
IRAS 17020+45  & 1782 & $9.344^{+4.020}_{-3.368} \times 10^{-3}$ & \\
Mrk 507        &14126 & $2.398^{+53.900}_{-1.653} \times 10^{-3}$& 43.62 (b)\\
IC 5063        &32205 & $9.072^{+4.476}_{-7.067} \times 10^{-3}$ & 42.87 (e)\\
Ark 564        &  888 & $8.482^{+2.099}_{-1.925} \times 10^{-3}$ & 43.38 (a)\\
\hline
3C 120		& 30600 & $1.929^{+1.530}_{-0.981} \times 10^{-4}$& 43.95 (d)\\
3C 390.3	&437915 & $3.675^{+2.267}_{-1.619} \times 10^{-3}$& 44.00 (d)\\
Ark 120		&237600 & $4.018^{+5.658}_{-2.411} \times 10^{-3}$& 43.88 (a)\\
IC 4329A	& 13510 & $4.058^{+2.302}_{-1.624} \times 10^{-4}$& 43.59 (a)\\
Mrk 509		& 26990 & $6.225^{+2.680}_{-1.989} \times 10^{-4}$& 44.03 (a)\\
NGC 3227	&319200 & $1.440^{+0.594}_{-0.465} \times 10^{-2}$& 41.66 (a)\\
NGC 3516	& 92475 & $5.320^{+2.147}_{-1.870} \times 10^{-3}$& 43.08 (a)\\
NGC 3783	& 29770 & $2.371^{+1.161}_{-0.715} \times 10^{-3}$& 42.90 (a)\\
NGC 4151	&102700 & $5.943^{+1.871}_{-1.484} \times 10^{-3}$& 42.62 (a)\\
NGC 4593	& 27140 & $6.578^{+13.53}_{-1.969} \times 10^{-3}$& 42.98 (a)\\
NGC 5548	& 34000 & $3.516^{+1.453}_{-1.161} \times 10^{-4}$& 43.41 (a)\\
NGC 7469	&127400 & $8.681^{+2.138}_{-1.853} \times 10^{-3}$& 43.25 (a)\\
F 9		& 40000 & $8.866^{+3.204}_{-3.206} \times 10^{-4}$& 43.91 (a)\\
\hline
F 303 		& 39936 & $(6.72 \pm 6.03) \times 10^{-3}$ & 43.03 (a)\\
PG 0026+129	& 39936 & $(1.31 \pm 1.92) \times 10^{-3}$ & 44.53 (a)\\
Mrk 586		& 39936 & $(2.57 \pm 0.75) \times 10^{-2}$ & 44.07 (a)\\
Mrk 1040	& 39936 & $(1.20 \pm 0.65) \times 10^{-2}$ & 42.40 (a)\\
NGC 985		& 39936 & $(3.47 \pm 1.76) \times 10^{-3}$ & 43.50 (a)\\
Mrk 279		& 39936 & $(2.32 \pm 0.84) \times 10^{-3}$ & 43.66 (a)\\
Mrk 841		& 39936 & $(1.14 \pm 0.93) \times 10^{-3}$ & 43.54 (a)\\
Mrk 290 	& 39936 & $(4.11 \pm 2.15) \times 10^{-3}$ & 43.22 (a)\\
\hline
\end{tabular}
\end{center}
First part of this table concerns the values of sources considered in
this paper. Second part shows the values obtained for BLS1 galaxies
(sample taken from Nikolajuk et al. 2006). 
Galaxies and their appropriate values shown in third part are taken
from O'Neill et al. 2005. Those all objects and several galaxies from
part one and two have been considered by Liu \& Zhang (2008) in their 
analysis.
%Therefore, we also add those objects from third part into our
%consideration.
Column~(1) lists the object name. Column~(2) shows the
arithmetic mean time of observations. Col~(3) gives the normalized
excess variance with errors. Those values have been calculated taking
mean $\langle T - 2 \Delta t \rangle$ and $\langle
\widetilde{A} \rangle$ as the product of those two quantities (see
equation~\ref{eq:variance_fin}).  The errors have been obtained from Monte
Carlo simulations of $\langle \widetilde{A} \rangle$ (see text).
Column~(4) shows logarithm of the luminosity in the 2-10 keV band. The
cosmology is $H_0 = 75\ \rm km s^{-1} Mpc^{-1}$ and $q_0 = 0.5$.
References in this column: (a) - O'Neill et al. 2005, (b) - Vaughan et
al. 1999, (c) - George et al. (2000), (d) - Merloni et al. 2003, (e) -
Bassani et al. (1999).
\end{table}
%%%%%%%%%%%%%%%%%%%%%%%%%%%%%%%%%%%%%%%%%%%%%%%%%%%%%%%%%%%%%%%%%%%%%%%

More accurate mass determination can be obtained if more measurements
are available for a given source. If a single but long observation is
available, it can be split into several parts.  Instead of direct
computation of an average excess variance we use a procedure which
allows for different duration of the lightcurve parts and higher
determination accuracy. The separate measurements are combined in the
following way: (i) we obtain the $k$th individual ratios:
\begin{eqnarray}
\widetilde{A}_k = {{(\sigma^2_{\rm nxs})}_k \over (T - 2 \Delta t)_k } \ ,
\end{eqnarray}
(ii) we fit this set of the individual $\widetilde{A}_k$ and calculate
the mean value of $\langle\widetilde{A}\rangle$ as weighted mean from
the minimum chi-square fit, in order to estimate the final value of
the X-ray excess variance
\begin{eqnarray}
\label{eq:variance_fin}
\sigma^2_{\rm nxs}= {\langle\widetilde{A}\rangle} 
{\langle T - 2 \Delta t\rangle}\ ,
\end{eqnarray}
where $\langle T - 2 \Delta t\rangle$ is the mean duration of a single
lightcurve.  The value of $\langle\widetilde{A}\rangle$ was also used
to calculate the black hole mass from the excess variance (Niko\l ajuk
et al. 2006)
\begin{eqnarray}
\label{eq:Mbhtotal}
\mbhvar= \frac{C}{\langle\widetilde{A}\rangle} \ .
\end{eqnarray}
Gierli\'nski et al. (2008) have shown that the method applies very
well to galactic sources at their hard states and to Seyfert 1
galaxies, and the proportionality constant, $C$, does not depend on
the source luminosity state.

The value of the scaling constant was determined to be $C = 0.96 \pm
0.02$ \Msun s$^{-1}$ (Nikolajuk et al. 2004), if based on the black
hole mass of Cyg X-1 equal to 10 \Msun\ (e.g.  Orosz 2003,
Herrero et al. 1995; Nowak et al. 1999; Gierli\'nski et
al. 1999). Nikolajuk et al.  (2006) obtained $1.92 \pm0.5$ \Msun\
s$^{-1}$, again based on Cyg X-1 scaling but taking the Cyg X-1 mass
of $20\pm5$ \Msun\ after Zi\'o{\l}kowski (2005). Recent study of
several galactic sources by Gierli\' nski et al. (2008) gave the value
$1.24\pm 0.06$\footnote{Note the difference in the definition of the
constant C between the present paper and Gierli\' nski et al. which
leads to a difference in units but not in the numerical value: $C_{\rm
Niko} = C_{\rm Gier} \times \nu_0^2$, where $\nu_0 = 1$ Hz. Hence, the
numerical value of the coefficients are the same in units \Msun\
s$^{-1}$ and \Msun\ Hz$^{-1}$.}. In the present paper we adopt the
value of 1.92 from Niko\l ajuk et al. (2006) for better consistency
with the previous study of BLS1 galaxies.

As it was mentioned by Nikolajuk et al. (2006), the accuracy of the
excess variance and $\mbhvar$ determination depend on the effect of
statistical error of variance measurement and power leaking from long
time-scales.  The errors are estimated by performing Monte Carlo
simulations for each source separately. The simulation generated a few
hundred sets of artificial data. From each lightcurve of one set was
calculated (\rms)$_k$, the coefficient $\widetilde{A}_k$ and, based on
the least squares method, the $\langle\widetilde{A}\rangle$.  Finally,
the 90\% error of $\langle\widetilde{A}\rangle$ was calculated due to
having a distribution of the hundreds values of them. The mass error
was determined from the error of $\langle\widetilde{A}\rangle$.

%%%%%%%%%%%%%%%%%%%%%%%%%%%%%%%%%%%%%%%%%%%%%%%%%%%%%%%%%%%%%%%%%%%%%%%
\begin{table}
\caption{The black hole masses.}
\label{tab:result}
\begin{center}
\begin{tabular}{l c c c c}%
\hline
Name & $\mbhlit$ & $\mbhvar$ & $\tr=\frac{\mbhlit}{\mbhvar}$ & \\
     &($10^{6}\ \rm M_{\odot}$) & ($10^{6}\ \rm M_{\odot}$)& \\
\hline
Mrk 335        & 14.2   & $1.00^{+1.24}_{-0.42}$    &$14.2^{+10.3}_{-7.9}$ \\
I Zw 1         & 33.4   & $1.55^{+9.82}_{-0.81}$    &$21.5^{+23.4}_{-18.5}$ \\
TON S180       & 22.5   & $1.21^{+0.31}_{-0.21}$    &$18.6^{+3.9}_{-3.8}$ \\
PHL 1092       & 293    & $0.40^{+6.50}_{-0.20}$    &$732^{+733}_{-690}$ \\
1H 0707+495    & 3.74   & $0.34^{+0.30}_{-0.12}$    &$11.0^{+6.0}_{-5.2}$ \\
Mrk 110        & 25.1   & $1.38^{+2.80}_{-0.67}$    &$18.2^{+16.7}_{-11.9}$ \\
Mrk 142        & 8.62   & $0.78^{+0.92}_{-0.35}$    &$11.0^{+9.0}_{-5.9}$ \\
Mrk 42         & 0.73   & $0.94^{+\infty}_{-0.54}$  &$0.78^{+0.89}_{-0.78}$ \\
NGC 4051       & 1.91   & $0.19^{+0.06}_{-0.04}$    &$10.1^{+2.6}_{-2.5}$ \\
%Mbh=(3 +2 -1)*10^5 from variability McHardy et al,2004,MNRAS
NGC 4395       & 0.36	& $0.062^{+0.032}_{-0.018}$ &$5.81^{+2.37}_{-1.98}$ \\
%Mbh=(0.05 +0.05 -0.04)*10^6 Vaughan et al. (2004)
PG 1211+143    & 146    & $5.3^{+21.8}_{-4.4}$      &$27.7^{+134.5}_{-22.3}$ \\
Mrk 766        & 1.52   & $0.45^{+0.65}_{-0.20}$    &$3.38^{+2.70}_{-2.00}$ \\
%Mbh=3.5*10^6 Botte et al. (2004);Mbh=3.47*10^6 O'Neill et al. (2005)
%Mbh~5*10^5 Vaughan & Fabian (2003) from break freq.
NGC 4258       & 39.0	& $9.45^{+5.75}_{-1.89}$    &$4.13^{+1.03}_{-1.56}$ \\
PG 1244+026    & 2.38   & $0.076^{+0.067}_{-0.030}$ &$31.3^{+20.4}_{-14.7}$ \\
MCG -6-30-15   & 2.84   & $1.49^{+0.43}_{-0.45}$    &$1.90^{+0.83}_{-0.42}$ \\
%Mbh=(2.9 +1.8 -1.6)*10^6 from variability McHardy et al. (2005)
PG 1404+226    & 12.1   & $0.51^{+0.92}_{-0.44}$    &$23.7^{+149.1}_{-15.2}$ \\
NGC 5506       & 88.0   & $5.11^{+2.20}_{-1.18}$    &$17.2^{+5.2}_{-5.2}$ \\
Mrk 478        & 40.2   & $1.55^{+2.21}_{-0.67}$    &$25.9^{+19.8}_{-15.2}$ \\
HB89 1557+272  & 4.17   & $1.88^{+2.56}_{-0.62}$    &$2.22^{+1.10}_{-1.28}$ \\
IRAS 17020+45  & 10.8   & $0.30^{+0.17}_{-0.09}$    &$36.0^{+15.4}_{-13.0}$ \\
Mrk 507        & 21.3   & $9.57^{+21.17}_{-9.14}$    &$2.23^{+47.30}_{-1.53}$ \\
IC 5063        & 55.0   & $3.67^{+12.94}_{-1.21}$   &$15.0^{+7.4}_{-11.7}$ \\
Ark 564        & 2.20   & $0.19^{+0.06}_{-0.04}$    &$11.6^{+3.1}_{-2.8}$ \\
\hline
\end{tabular}
\end{center}
Column~(1) lists the object name. Column~(2) shows the masses of black
holes without errors taken from Table~\ref{tab:Optdata}. Column~(3)
shows the black holes masses obtained from the variance method. The
ratios of the two masses are given in Column~(4).
\end{table}
%%%%%%%%%%%%%%%%%%%%%%%%%%%%%%%%%%%%%%%%%%%%%%%%%%%%%%%%%%%%%%%%%%%%%%%

%<<<<<<<<<<<<<<<<<<<<<<<<<<<<<<<<<<<<<<<<<<<<<<<<<<<<<<<<<<<<<<<<<<<<<<<<<
\section{Results}
\label{sect:results}

%--------------------------------------------------------------------
%\subsection{Mass measurements}

Since NLS1 galaxies are known to be relatively more variable than BLS1
galaxies (e.g. Hayashida 2000, Czerny et al. 2001; McHardy et
al. 2004; Papadakis 2004), we do not expect the standard X-ray excess
variance method of mass measurement, with the same value of the constant C,
 to work for these objects.

Instead, we search here for the discrepancies between variability
based method and other methods of mass measurements, and we analyse
the dependence of these discrepancies on the source properties.

\begin{figure}
\epsfxsize=8.8cm
\epsfbox[80 200 580 700]{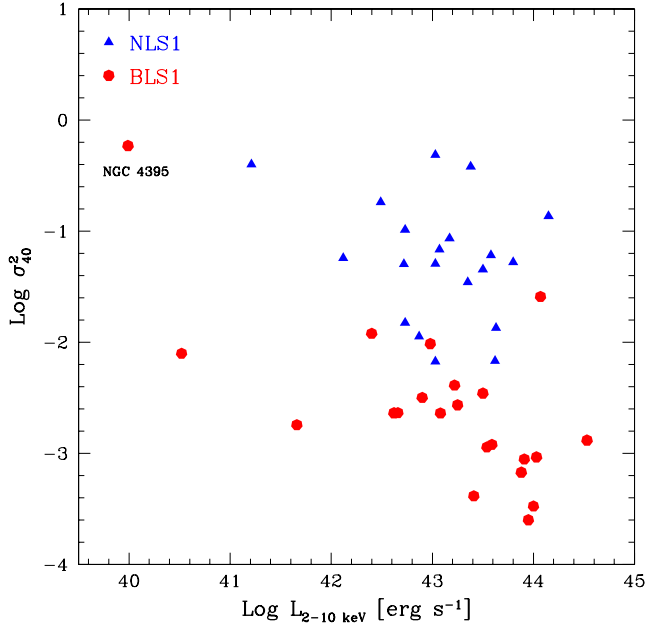}
\caption{The renormalized excess variance against the X-ray luminosity
in the 2-10 keV range. The renormalized excess variance $\sigma^2_{40}
=$ \rms$\times (40000/\langle T \rangle)$ i.e. to the value expected
for the duration of the lightcurve equal to 40 000 s.  The whole
sample of galaxies is presented in Table~\ref{tab:TsigLx}.  We enlarge
our sample by a few galaxies taken from 0'Neill et al. (2005) in order
to be consistent with sample of Liu \& Zhang (2008).  }
\label{fig:sig-Lx}
\end{figure}
\begin{figure}
\epsfxsize=8.8cm
\epsfbox[80 200 580 700]{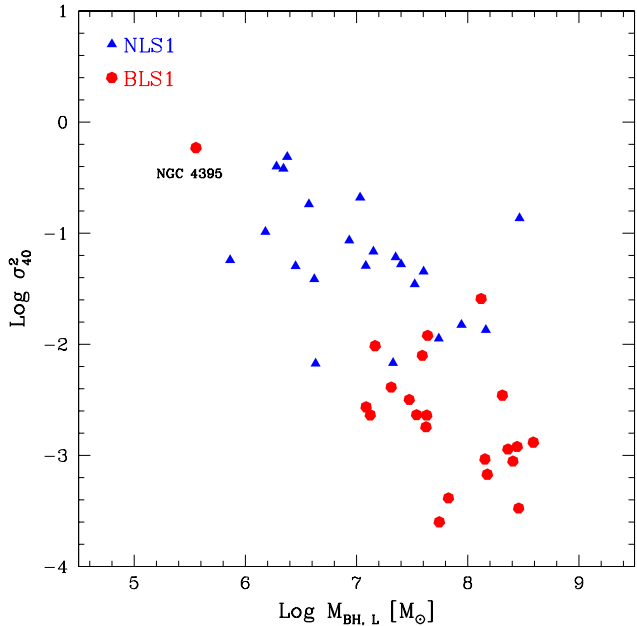}
\caption{The same as in Fig.~\ref{fig:sig-Lx} but versus the black
hole masses taken from optical determination. 
%The values of $\mbhlit$ were taken from O'Neill et al. (2005).
}
\label{fig:sig-M}
\end{figure}

The details on the lightcurves analysed for the purpose of
this work are given in Table~\ref{tab:Xdata} and the values of the
X-ray excess variance for a whole sample of both NLS1 and BLS1
galaxies are given in Table~\ref{tab:TsigLx}.

We first check single parameter relations between the X-ray excess
variance and the black hole mass (taken from optical determination)
and the X-ray luminosity. We renormalize the excess variance to the
value expected for the duration of the lightcurve equal to 40 000 s,
i.e. $\sigma^2_{40} =$ \rms\ $\times (40000/\langle T \rangle)$. The
results are plotted in Figs.~\ref{fig:sig-Lx} and \ref{fig:sig-M}.

There is much better correlation with mass than with luminosity in the
whole combined sample of NLS1 and BLS1 galaxies (correlation
coefficient equal to -0.69 and -0.39 correspondingly). However, the
correlation with the luminosity is basically driven by the BLS1
objects. If only NLS1 are selected, the correlation of
the excess variance with the black hole mass is stronger than
with the X-ray luminosity (-0.44 and -0.27 correspondingly).

The correlation with the mass suggest that the X-ray excess variance
can possibly be used for mass determination in NLS1 galaxies. We
therefore calculate the masses of the black holes in our sample using
the equation~(\ref{eq:Mbhtotal}).

The values of the black hole masses derived from the standard excess
variance, $\mbhvar$, the values of the masses based on other methods
(reverberation, stellar velocity dispersion, etc),
$\mbhlit$, and the ratio of the two values for each source in our
sample of NLS1 galaxies, $\tr=\frac{\mbhlit}{\mbhvar}$, are given in
Table~\ref{tab:result}.

\begin{figure}
\epsfxsize=8.8cm
\epsfbox[80 200 580 700]{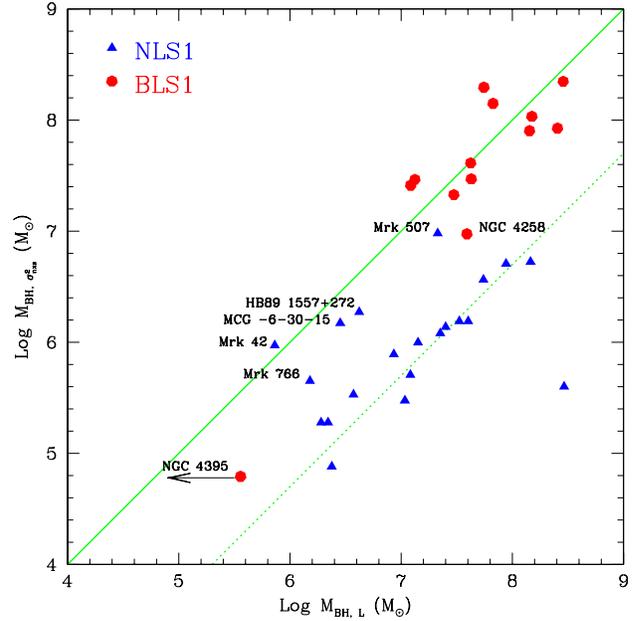}
\caption{ Black hole masses estimated from X-ray variance 
method,$\mbhvar$, versus masses taken from literature, $\mbhlit$. The
continuous line shows the relation $\mbhlit = \mbhvar$. The dot line
shows the relation $\mbhlit = 20\times \mbhvar$.
We can see that small subset of NLS1 objects which
do not require any rescaling like BLS1 galaxies. The arrow close to NGC
4395 indicates that $\mbhlit$ of those galaxy may be smaller in
agreement with Filippenko \& Ho (2003) ($\mbhlit \sim 10^4-10^5$\Msun).}
\label{fig:Mbh}
\end{figure}

In Fig.~\ref{fig:Mbh} we show the plot of the mass from variability
vs. mass from the literature for our sample of galaxies and for BLS1
galaxies from Nikolajuk et al. (2006). The BLS1 sources cluster along
the continuous line marking the ratio~1. Apart from two sources, the
value of $\tr$ for all objects is within the limits 0.3 - 3. The
source with the smallest mass, NGC 4395 is shifted from the line ($\tr
= 5.81$ for this source). We have to note that our obtained value
($6.2 \times 10^4$\Msun) is similar to other values obtained from
other methods based on X-ray variability: $\sim 6 \times 10^4$\Msun\
(Vaughan et al. 2005; $M_{\rm BH}-P(\nu)\times \nu=10^{-3}$ method);
$10^4$-$10^5$\Msun\ (Shih et al. 2003; $M_{\rm BH}- \nu_{\rm br}$
method).  Other black hole mass determinations for this source also
gave lower values than the direct reverberation (photoionization,
Kraemer et al. 1999; $\Rblr$ scaling with luminosity, Filippenko \& Ho
2003) have mentioned \mbh\ in NGC 4395 of order of $\sim 10^4$-$10^5$
\Msun. In such case the ratio $\tr$ would be equal to 0.6-1.61,
respectively. On the other hand, the upper limit from stellar
dispersion velocity in the central 3.9 pc ($6.2 \times 10^6
M_{\odot}$, Filippenko \& Ho 2003) is still consistent with the
reverberation value. In the case of NGC~4258 the BH mass is estimated
precisely from maser emission method. Our determination gives the
upper limit of \mbh\ $15.2 \times 10^6$ \Msun\ and hence the lower
limit of $\tr = 2.5$, so the agreement is not quite satisfactory.

\begin{figure}
\epsfxsize=8.8cm
%\epsfbox[80 200 580 700]{MNik_fig4.eps}
\epsfbox[70 100 560 585]{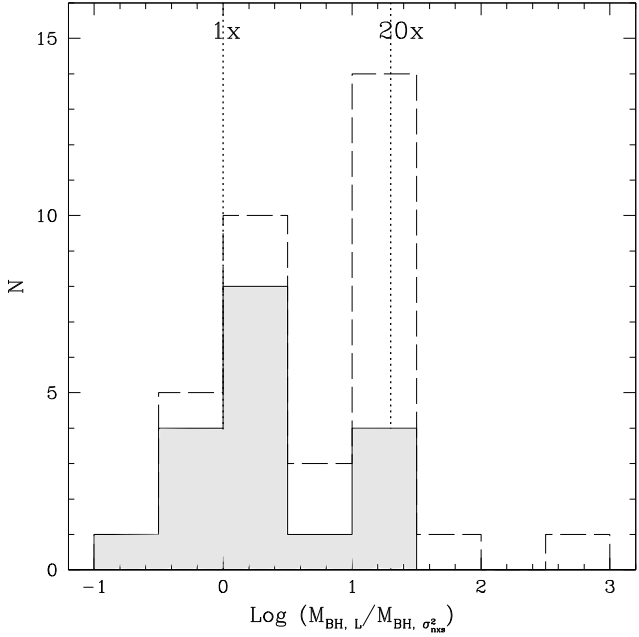}
\caption{
Two histograms showing the number of our sources (NLS1 + BLS1) in a
particular bin of logarithm of the mass ratio $\mbhlit/\mbhvar$.
Histogram, which is drawn by long dash line, concerns all sources in
our sample. Shaded histogram, drawn by continuous line concerns
objects with masses $\mbhlit$ estimated based on the reverberation
method or by maser method (only NGC 4258). We can see bimodality in
both cases.
}
\label{fig:histo.MM}
\end{figure}

The NLS1 sources show larger dispersion than BLS1 sources. Most of them
cluster along the dotted line, marking the ratio of 20. 
This simply reflects the fact that in both plots of \rms\
(Figs.~\ref{fig:sig-Lx} and \ref{fig:sig-M}) the NLS1 objects lie on average
higher, as expected.

The existence of two characteristic values of the ratio is already
suggested by Fig.~\ref{fig:Mbh} but it is seen more clearly from the
direct plot of the mass ratio. In Fig.~\ref{fig:histo.MM} we present
the results on the variability level in our sample in a form of two
histograms showing the number of sources in a particular bin of the
$\mbhlit$ and $\mbhvar$ mass ratio. First histogram (drawn by
long-dashed line) considers all objects in our sample of NLS1 and BLS1
galaxies.  Second histogram (drawn by continuous line and shaded)
shows galaxies which $\mbhlit$ are taken from the reverberation or the
maser method (for NGC 4258) only. Both histograms are similar.  The
distributions show two separate peaks, one close to unity, and one
close to the factor 20. BLS1 galaxies occupy the first peak while most
of NLS1 galaxies make the second one.

Our data come from two different instruments (ASCA and Rossi-XTE; see
Table~\ref{tab:Xdata}) which cover slightly different energy range we
checked whether this may have a strong impact onto our results. For
MCG -6-30-15 we had both ASCA and Rossi-XTE data, so we calculated the
X-ray excess variance for this source independently from both
instruments. The ASCA variance was by a factor 1.957 higher than
Rossi-XTE variance which is consistent with ASCA being sensitive in
softer energy band and with the general trend of lower normalization
of the power spectrum in higher energies (e.g. Uttley \& McHardy
2005). We then made an experiment and multiplied all variance values
from Rossi-XTE by this factor and replotted the histogram. It still
did not change and showed apparent bimodality.

However, there exists a small subset of our NLS1 sources which does
not require that multiplication and the value of $\tr$ is the same
like in BLS1 cases (MCG -6-30-15, Mrk 42, Mrk 507, Mrk 766 and quasar
HB89 1557+272).

Since the behaviour of this small subset of NLS1 is puzzling and may
help to understand the origin of the variability we analyse the
possible cause of this behaviour.

\begin{figure}
\epsfxsize=8.8cm
\epsfbox[80 200 580 700]{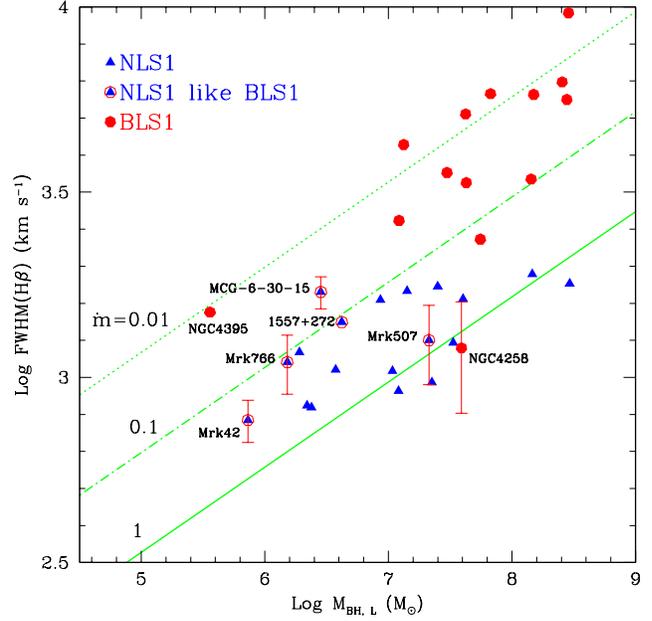}
\caption{The Full Width at Half Maximum of \Hb\ line produced in 
Broad Line Region in NLS1 galaxies versus black hole masses
$\mbhlit$. The continuous, dot-short-dash and dot lines represent
accretion rate in Eddington units, $\dot m$ (respectively 1, 0.1 and
0.01).  }
\label{fig:fwhm}
\end{figure}

The location of these slowly variable NLS1 on the FWHM(H$\beta$)
versus the mass diagram is shown in Fig.~\ref{fig:fwhm}. The errors
are large but the sources do not seem to be located differently from
the other NLS1 objects. In Fig.~\ref{fig:fwhm} we also indicate the
lines corresponding to the fixed dimensionless accretion rate, or the
Eddington ratio, $\mdot = \dot M/\dot M_{\rm Edd}$.  We have
calculated $\mdot$ using equation taken from Czerny, R\'o\.za\'nska \&
Kuraszkiewicz (2004): ${\rm FWHM} = A M_{\rm BH}^{0.15}
\mdot^{-0.35}$, where $A$ is constant. They derived this formula using
Kaspi et al. (2000) relationship for $\Rblr$ and assuming the
geometrical factor $f = \sqrt{3}/2$.  We have corrected Czerny et
al. formula by taking new estimated by Bentz et al. (2007)
relationship $\Rblr \propto [\lambda L_{\lambda} (5100 \rm
\AA)]^{0.54}$ and assuming $f = \sqrt{5.5}/2$. 
Therefore, the corrected formula using by us in this paper is:
\begin{eqnarray}
\label{eq:dotm}
\dot m = \Big[\frac{\rm FWHM(H\beta)}{1650 \ \rm kms^{-1}}\Big]^{-3.70}
\Big(\frac{M_{\rm BH}}{10^8 \rm M_{\odot}}\Big)^{0.85} \ .
\end{eqnarray}
In Fig.~\ref{fig:fwhm} the division into BLS1 and NLS1 objects
corresponds roughly to the division into $\dot m < 0.1$ and $\dot m >
0.1$ for a $10^7 M_{\odot}$ black hole.  For other values black hole
masses the exact classification according to the spectra (NLS1 versus
BLS1) and according to the Eddington ratio do not overlap. The issue
of the usefulness of the current definition of NLS1 was already
raised by Mathur \&\ Grupe (2005a,b) in the context of comparison of
optically-selected and X-ray selected NLS1.

\begin{figure}
\epsfxsize=8.8cm
\epsfbox[80 200 580 700]{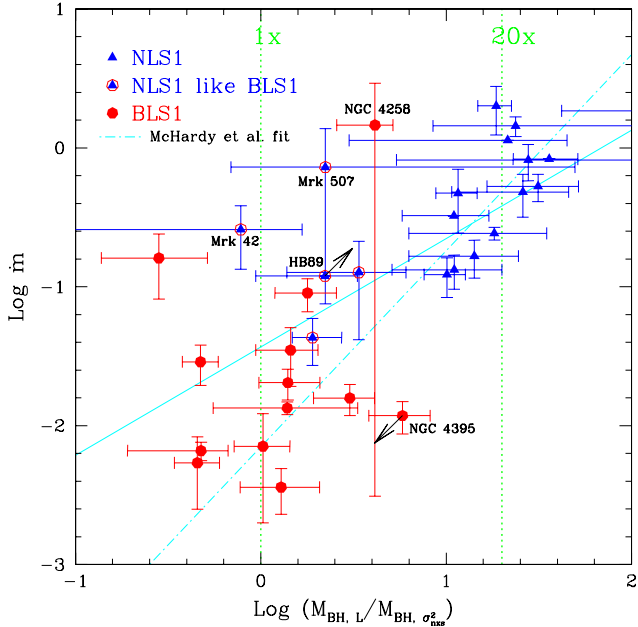}
\caption{
The dimensionless accretion rate, $\dot m$, plotted versus the ratio
$\mbhlit/\mbhvar$. The dotted vertical lines show how many times
should be multiplicated $\mbhvar$ in order to obtain $\mbhlit$.  The
continuous line shows the best fit to our sample of galaxies. The
dot-short dash line is the McHardy et al. fit. 
See Sect.~\ref{sect:multi} for details.
}
\label{fig:mdotMM}
\end{figure}

In Fig.~\ref{fig:mdotMM} the NLS1 objects which have slower
variability than most NLS1s are naturally located at the mass ratio
$\sim 1$, but their Eddington ratios span rather wide range. This may
be related to relatively large errors for a single object. 

The formula used by us derives the value of the dimensionless
accretion rate partially from the black hole mass, and relies on the
scaling of the Broad Line region so it is not independent from the
mass ratio in the statistical sense. Therefore, the result may be
biased by the underlying correlation. However, at present this is the
simplest way to see the trend with the Eddington ratio. We will
discuss Fig.~\ref{fig:mdotMM} again in Sect.~\ref{sect:multi}.

In order to gain additional insight which may not be biased by
determination of $\dot m$, we consider the dependence of the
variability on the soft X-ray slope.

\begin{figure}
\epsfxsize=8.8cm
\epsfbox[80 200 580 700]{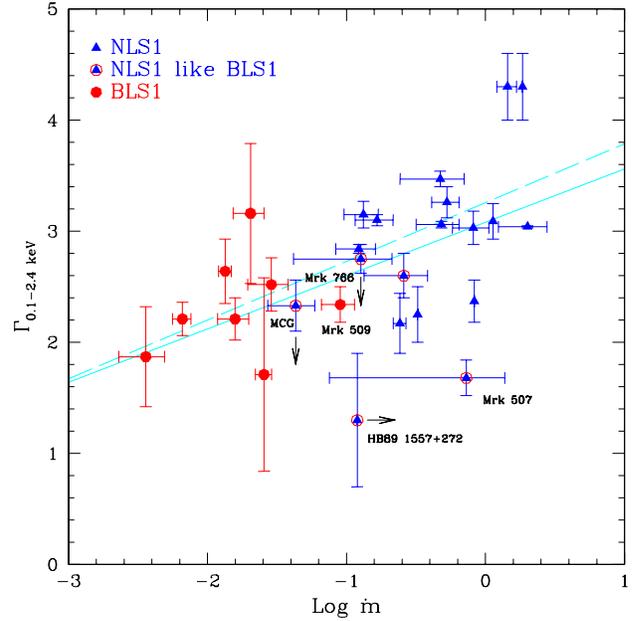}
\caption{The soft X-ray slope, \Gs, plotted versus
the dimensionless accretion rate, $\dot m$.
The continuous line shows fit to our sample of galaxies.
The long-dash line shows fit to the same sample subtracted by the
subset of 5 'NLS1 like BLS1' objects.
}
\label{fig:mdotG2}
\end{figure}

The spectral slope in 0.1-2.4 keV range in our sample generally
correlates with the dimensionless accretion rate (see
Fig.~\ref{fig:mdotG2}). For two slowly variable NLS1 the X-ray slope
is flatter than expected from the overall correlation while for other
three sources are close to the best fit line. 
The continuous line represents our best fit to our sample
(correlation coefficient of this fit is 0.515): $\Gamma_{0.1-2.4 \rm
keV} = (0.48 \pm 0.16)$ log $\dot m + (3.08 \pm 0.18)$.  The long-dash
line shows fit to the same sample but without the subset NLS1
like BLS1 galaxies. This fit is $\Gamma_{0.1-2.4 \rm keV} = (0.53 \pm
0.14)$log $\dot m + (3.26 \pm 0.16)$ with correlation coefficient
0.650.

\begin{figure}
\epsfxsize=8.8cm
\epsfbox[80 200 580 700]{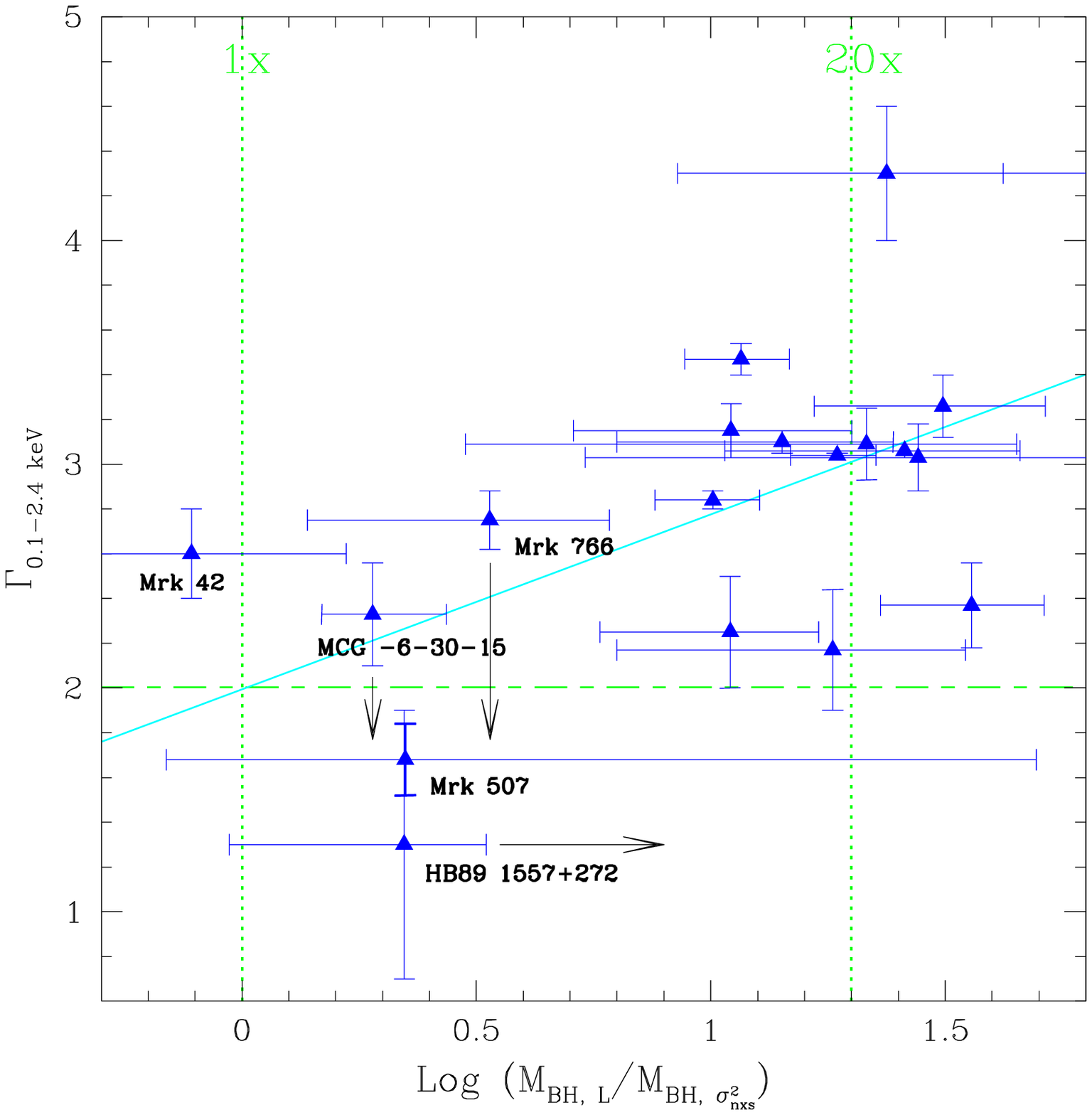}
\caption{The soft X-ray photon index, \Gs, plotted versus the ratio 
$\mbhlit/\mbhvar$ for NLS1 galaxies. The dotted vertical lines
represent the same as in the Fig.~\ref{fig:mdotMM}. The dashed line
shows arbitrary border \Gs\ = 2. Horizontal arrow near HB89 indicates
that the BH mass in this quasar may be underestimated. Vertical arrows
mark that in agreement with Branduardi-Raymont et al. (2001) or Lee et
al. (2001) the values of \Gs\ in MCG -6-30-15 and Mrk 766 may be
lower. Continuous line represents the best fit \Gs $=(0.78 \pm 0.21)$
log$(\mbhlit/\mbhvar) + (2.00\pm 0.27)$ with correlation coefficient
equal to 0.668. }
\label{fig:Gamma}
\end{figure}

Now we present the relation between the soft X-ray slope and the mass
ratios for our sample of NLS1 galaxies (see Fig.~\ref{fig:Gamma}). A
strong trend with a soft X-ray slope is seen again. Since the
measurement of the soft X-ray slope is more direct than the estimate
of the accretion rate or bolometric luminosity, this result indeed
supports the claim of McHardy et al. (2006) that the black hole mass
alone does not determine the variability properties if we include
objects with steep soft X-ray spectra into our sample. The change of
variability properties with the change of the spectral slope is not
surprising, since the underlying mechanism may be different as well.

If indeed there are two separate mechanisms of the hot plasma
formation, separate for soft spectra and separate for hard X-ray power
law, then the dependence of variability level on the sources
properties may be rather bimodal, reflecting a rapid state
transition, instead of showing a continuous dependence on the
Eddington ratio, as proposed by McHardy et al.

Our sample may not still be large but at present the
distribution looks rather bimodal, with sources concentrating around
the ratios $\sim 1$ and $\sim 20$ instead of continuously covering the
whole diagram between some minimum and maximum values, as it would be
expected from a continuous dependence on $\dot m$. The corresponding
histogram of $\dot m$ does not show such a bimodality. However, a
histogram of slopes in our sample again shows a bimodal behaviour,
with objects clumping around $\Gamma \sim 2$ and $\Gamma \sim
3$. This supports the view that indeed a spectral change from hard to
soft X-ray slope is underlying the change in variability properties of
AGN.

\subsection{Multidimensional fits}
\label{sect:multi}

In the previous section we concentrated on the relation between the
X-ray excess variance with the black hole mass but there were
suggestions that the variability may correlate both with the mass and
one more parameter like X-ray luminosity or bolometric luminosity.

Since the division of objects according to the Eddington ratio seems
to be based on better physical grounds we test whether the dependence
of the variability level on this ratio can account for the issue of
enhanced NLS1 variability. Such an idea was put forward by McHardy et
al. (2006) for a small sample of AGN.

The dependence of the mass ratio, or the variability enhancement, on
the dimensionless accretion rate for our sample is shown in
Fig.~\ref{fig:mdotMM}. 
We have calculated $\dot m$ using equation~(\ref{eq:dotm}) with the data of
\fwhm\ and $\mbhlit$ taken from
Tables~\ref{tab:tab1}-\ref{tab:Optdata}.  
Our best fit (solid line in figure) to the whole
sample (without NGC 5506 i IC 5063 for which we have no $\dot m$
estimate) gives
\begin{equation}
\log \dot m = a \log \tr + b ,
\end{equation} 
with $a = 0.79 \pm 0.18$ and $b = -1.43 \pm 0.14$. The correlation is
statistically significant (correlation coefficient 0.712).

We can compare this result with the result of McHardy et
al. (2006). If we assume that the high frequency slope of the power
spectrum is equal 2, and the normalization of the frequency multiplied
power spectrum is independent from the object we can relate their
result based on frequency break to our result involving the high
frequency tail. With this assumption, the result of McHardy et
al. (2006) imply we should expect a linear trend with the slope $\sim
1$, or more precisely, a relation
\begin{equation}
\log \dot m = {A - B \over B} \log \tr + const,
\label{eq:mdotr}
\end{equation} 
where $A = 2.17^{+0.32}_{-0.25}$ and $B=0.90^{+0.3}_{-0.2}$ are the
coefficients from the McHardy et al. fits. Our larger sample is
roughly consistent with this result since our slope $a = 0.79$ and the
slope implied by McHardy et al. (2006) results is $a = $1.4. Since the
dependence on the dimensionless accretion rate implied by the results
based on the frequency break is even steeper it means that the overall
trend of an increase of the mass discrepancy between the two methods
with the Eddington ratio seems to be real although determination of
this ratio is highly uncertain.

We also show in this figure the fit calculated by McHardy 
et al. (2006) made for 10 objects (i.e. 5 BLS1 and 5 NLS1).
%Authors selected 10 objects i.e. 5 BLS1 and 5 NLS1.
%NGC~3227,3516,3783,4151,4395,4051,5506, MCG -6-30-15, Mkn 766, Akn 564.
This fit is represented by the dot-short-dash line with the slope $(A
- B)/B = 1.41$ and $const$ value chosen in arbitrary way. Close to
this fit lie our second fit (not shown in Fig~\ref{fig:mdotMM}).  This
fit is calculated for our sample of the same objects as in McHardy et
al., but without NGC 5506.  The value of slope $a$ of our second fit
($a =1.07\pm 0.35$) and correlation coefficient (0.780) indicates that
our approach is indeed equivalent to McHardy et al. (2006) analysis.

However, such a two-dimensional fit simply joints the extreme values
of our mass ratios in Fig.~\ref{fig:mdotG2} and does not provide an
explanation of the bimodal distribution seen in
Fig.~\ref{fig:histo.MM}.

We also repeat the two-dimensional fit to both the black mass and the 
X-ray luminosity proposed by Liu \& Zhang (2008). If we allow for a general 
relation between the X-ray excess variance in our sample of 43 objects 
(renormalized to 40 000 s lightcurves) and the X-ray luminosity
\begin{equation}
\log \sigma^2_{\rm nxs} = A \log \mbhlit + B \log L_{\rm 2-10 keV} + C
\end{equation}
we obtain the fit parameters $A = -0.84^{+0.23}_{-0.20}$, $B =
-0.0088^{+0.039}_{-0.039}$, $C = 4.5^{+1.7}_{-1.6}$. This sample
includes the objects with the mass taken from the scaling relation
since the number of sources with direct reverberation mass measurement
is low. However, it seems that the most important effect is the
careful measurement of the X-ray excess variance. If we take only the
excess variance measurements obtained by O'Neill et al. (2005) and
used by Liu \& Zhang, and supplement them with our measurements only
for sources which were not included in their study, the result is
widely different: the dependence on the black hole mass is weak ($A =
-0.058$) while the dependence on the X-ray luminosity is strong ($B =
-0.36$) irrespectively of the use of scaled values of the black hole
mass. Our measurements of the X-ray excess variance for a number of
objects differ from the values obtained by O'Neil et al. (2005) due to
another way of averaging and different length of single curves.

\section{Discussion}
\label{sect:discussion}

The X-ray variability of the BLS1 galaxies at high frequencies depends
only on the black hole mass, and the same is true for the galactic
sources in their hard states (Gierli\'nski et al. 2008). Therefore,
the X-ray excess variance is a viable method of black hole mass
determination (e.g. Papadakis 2004; Nikolajuk et al. 2004; Awaki et
al. 2005; Zhang et al. 2005; Markowitz et al. 2006; Nikolajuk et
al. 2006; see also O'Neill et al. 2005).

The issue complicates when the sources with soft X-ray spectra are
considered. In the case of AGN the sources with the soft X-ray spectra
usually belong to the NLS1 galaxy class, and usually they are more
variable than the corresponding broad line objects
(Iwasawa et al. 2000; Markowitz \& Edelson 2004; 
Vaughan et al. 2005). McHardy et al. (2006) suggested that 
this effect should be taken into account
in a form of continuous dependence of the variability rate not just on
the black hole mass but also on the Eddington ratio. They have found a
relation which roughly represents the variability properties in a
combined sample of NLS1 and BLS1 objects.

In the present paper we studied a larger sample of 21 NLS1
galaxies. We confirm the dependence of the variability properties on
the accretion rate if NLS1 galaxies are included. However, our study
indicated that such a parametrization of variability as proposed by
McHardy et al. (2006) for AGN may not fully reflect the complexity of the
issue.

We have found that the essential effect is not in the dependence on
the accretion rate but likely in the dependence on the soft X-ray
slope. The slope is apparently related to the character of
variability, as is seen for example in the study of the
energy-dependent fractional variability by Gierli\'nski \& Done
(2006). Typical soft NLS1 like NGC 4051 showed much higher amplitude at
lower energies, below 0.5 keV than at 10 keV, indicating a pivoting
trend while the hard spectrum NLS1, MCG -6-30-15 showed the same
amplitude at very low and very high energy, i.e. no change of the
spectral slope. What is more, we see an indication of a bimodal
behaviour connected with predominantly bimodal distribution of the
soft X-ray slopes of AGN in our sample while the distribution of the
accretion rates seems to be continuous. This explains why Gierli\'nski
et al. (2008) do not find the dependence of the variability rate on
luminosity in their study of the galactic sources: the sources were
selected to represent the hard state only and the evolutionary stages
with soft spectra were not selected for consideration.

The fact that it is not the luminosity change but the spectra change,
which affects the fastest variability, hints for a different, or
additional variability mechanism when the spectrum becomes softer.

We would like to emphasise the fact that our approach in calculating
the normalized excess variance, based on computation weighted mean
\rms\ from the minimum chi-square fit, is more reliable and gives more
accurate results than direct computation. Indeed, the way of
calculating the excess variances is very important and it influences
on obtained results. The findings depends on it very strongly.  From
this follows differences between our and Liu \& Zhang (2008)
results. When we repeated calculation of Liu \& Zhang taking only
their values of \rms, we obtained results similar to them.

Nevertheless, we must note that we had still too small sample of
galaxies, the data came from two different instrument (ASCA and
Rossi-XTE), and particularly the number of sources with mass
determination from monitoring was far from satisfactory.  In order to
resolve the fact that existence of bimodal distribution is realised
in nature or not we need enlarge sample, especially by NLS1 objects
which behave in the X-ray variance like BLS1 galaxies.

%%%%%%%%%%%%%%%%%%%%%%%%%%%%%%%%%%%%%%%%%%%%%%%%%%%%%%%%%%%%%%%%%%%%%%%
\section*{Acknowledgments}

We would like to thank Yuang Wei-Min for useful discussions.  The
present work was supported by grants 1P03D00829 of the Polish State
Committee for Scientific Research and by the Polish Astroparticle
Network 621/E-78/SN-0068/2007. We also acknowledge the use of data
obtained through the HEASARC online service provided by NASA/GSFC.

\ \\
This paper has been typeset from a \TeX/\LaTeX\ file prepared by the
author.

\end{document}